# Can AI decrypt fashion jargon for you?

**Yuan Shen**
University of Illinois at
Urbana-Champaign
Champaign, IL 61820, USA
yshen47@illinois.edu

**Shanduojiao Jiang**
University of Illinois at
Urbana-Champaign
Champaign, IL 61820, USA
sj10@illinois.edu

**Muhammad Rizky Wellyanto**
University of Illinois at
Urbana-Champaign
Champaign, IL 61820, USA
wellyan2@illinois.edu

**Ranjitha Kumar**
University of Illinois at
Urbana-Champaign
Champaign, IL 61820, USA
ranjitha@illinois.edu

## Abstract
When people talk about fashion, they care about the underlying meaning of fashion concepts (e.g., style). For example, people ask questions like what features make this dress smart. However, the product descriptions in today's fashion websites are full of domain-specific and low-level words. It is not clear to people how exactly those low-level descriptions can contribute to a style or any high-level fashion concept.

In this paper, we proposed a data-driven solution to address this concept- understanding issue by leveraging a large number of existing product data on fashion sites. We first collected and categorized 1546 fashion keywords into 5 different fashion categories. Then, we collected a new fashion product dataset with 853,056 products in total. Finally, we trained a deep learning model that can explicitly predict and explain high-level fashion concepts in a product image with its low-level and domain-specific fashion features.

## Author Keywords
fashion; deep learning; code switching

## CCS Concepts
•**Human-centered computing** → **Human computer interaction (HCI)**; •**Computing methodologies** → **Computer vision;**

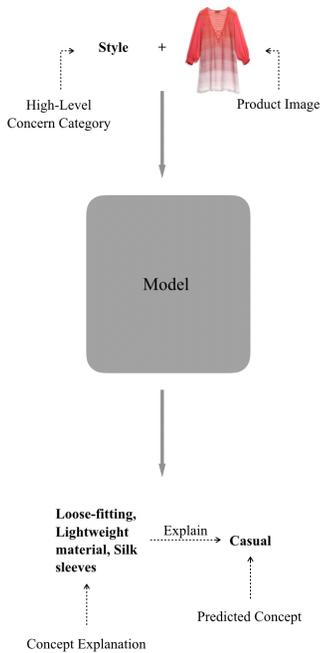

**Figure 1:** Model Application

## Introduction

*"If I read that a square-necked, white silk sweater is very smart, it is impossible for me to say, ..., which of these four features (sweater, silk, white, square neck) act as signifiers for the concept 'smart'?"*

– Roland Barthes in *The Language of Fashion*

When people talk about fashion, they are curious about the underlying meanings of fashion concepts, like style, with low-level words. However, in today's fashion websites, it is easy to get lost in the long and domain-specific descriptions. For instance, e-commerce sites prefer product name like "Etoile Sequin Fringe Dress" over "Party Dress". It is not clear to people how exactly those low-level descriptions can contribute to a style or any high-level fashion concept. We are wondering if we can build a data-driven solution to solve this concept understanding problem. The proposed data-driven solution could empower ordinary people to understand their high-level fashion concept with explicit low-level attributes.

Existing research in the Computer Vision community explored a few vision-based or multi-modal models on fashion product representation. In particular, Feng et al. [2] introduced interpretable attribute embedding, which was used in a fashion outfit composition task. Our model is built on top of their attribute product representation. However, instead of outfit composition, we leverage the attribute embedding set to help people reason why a product fits a high-level concept.

In this paper, we proposed a deep learning solution that explicitly connects objective attributes or domain-specific words with high-level fashion concepts in our model architecture. We first collected and defined 1546 fashion keywords grouped into 5 different categories. And then we crawled 4 fashion websites with 853,056 products in total. During model training, we adopt a two-step training procedure. We first separately trained three different fashion attribute encoders on color, shape, and pattern. Later, we represented each product with the 3 attribute embeddings. Finally, the model will predict high-level concepts based on low-level attribute embeddings.

The model enables the following interaction. Given a product image and a high-level concern category (like an event, time or location), the model can not only predict the corresponding high-level concept but also explain that high-level concept with low-level words.

In summary, the key contributions are as follows:

- proposed the concept understanding issue in the fashion e-commerce product descriptions
- built a deep learning model that can use text or visual low-level features to explain high-level concepts

## Related Work

Explainable AI has been explored from different perspectives. Olah et al. extracted intermediate visual hidden layers to find meaningful semantic representation [4]. Lakkaraju et al and Ribeiro et al. focuses on trust issues between people and AI models by introducing interpretability [5]. Specifically in the fashion domain, Feng et al. built interpretable semantic layers to improve outfit compatibility prediction tasks [2] by introducing mutual independence loss. We find it necessary to explain hard and professional concepts with simple and common ones. Feldman reported a discovery that complex concepts are hard to be learned by people [1]. Finally, we found Amazon Fashion already deployed occasion-based product descriptions for fashion. However, they did not explain how they come up with the proposed

| Model | Metric | Result |
|---|---|---|
| $Enet_{color}$ | MSE | 0.142 |
| $Enet_{shape}$ | MRR | 0.0167 |
| $Enet_{pattern}$ | MRR | 0.0478 |

**Figure 3:** Attribute Encoder Performance. MSE is mean square error; MRR is mean reciprocal rank, widely used in information retrieval.

| Product Embedding | MRR (high-level concepts) |
|---|---|
| Attribute Embeddings | 0.0095 |
| ResNet50 Embeddings | 0.0091 |

**Figure 4:** Ablation study of product embedding. By substituting the resNet50 embedding with attribute embedding set of color, shape and pattern, we observed a 3.3% performance increase in terms of MRR

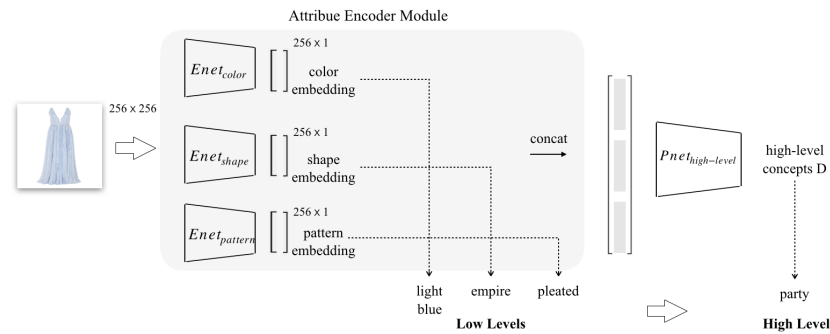

| Model | Architecture |
|---|---|
| $Enet_{color}$ | CD32 + CD64 + CD128 + Flatten + FC256 + FC15 + Sigmoid |
| $Enet_{shape}$ | Unet encoder + FC256 |
| $Enet_{pattern}$ | ResNet-50 + CD512 + CD256 + Softmax |
| $Pnet_{high-level}$ | CD512 + CD256 + Softmax |

**Figure 2:** Model Architecture : low-level (objective words) embedding is trained first, then the output will become the input of the high-level (e.g. scenario) prediction model. We used top 5 RGB colors as labels for the color encoder. For $Enet_{color}$, we added an additional linear layer to project from 15 to 255 dimensions. For pattern, we first converted the original image to grayscale and then feed into $Enet_{pattern}$. CD[x] stands for the convolution layer with output feature dim x which followed by BatchNorm, ReLU, and MaxPooling. We self-trained our segmentation model with the fashion clothes dataset using Unet [6] and extracted color labels with [3].

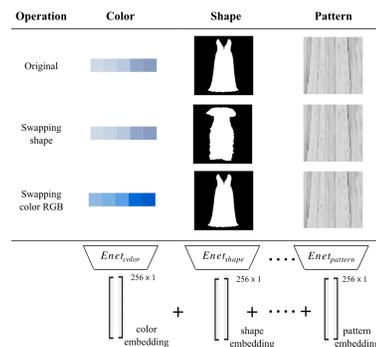

**Figure 5:** *AttributeSwap*. During inference time, we can increase our data variation to draw insights about concepts by swapping individual attribute embedding. For example, by swapping the original pattern with a camouflage pattern, we synthesize a new camouflage y-line dress, which may not exist in our original dataset.

occasions, whereas we connect low-level and high-level concepts both explicitly and transparently.

### Data
First, we collected our fashion lexicon, where each vocab is manually labeled into 5 categories. In total, we have 375 materials, 65 patterns, 270 shapes, 556 styles, and 280 trims.

Then we crawled 4 fashion websites (Farfetch, Amazon, Zara and Moda Operandi) with 853,056 products in total. For each product, we recorded data including name, description, image, etc. To get product attribute labels, we extracted fashion concepts from the product description, by checking whether it contains any fashion vocab in the lexicon defined above.

### Model
The architecture has two training steps. Since our goal is to provide insights about concept understanding, we adopt the simplest network design in our module. First, in the Attribute Encoder Module (AEM), it will generate a product embedding from the input image. Then, we will predict the high-level concepts in the next step.

To note in the AEM, instead of using image representation generated by the pre-trained ImageNet architectures, the model uses separately-trained attribute embeddings (color, shape, and pattern) to represent the product altogether. One of the advantages of using this attribute embedding set representation is that it improves the performance of high-level concept prediction (Figure 4). On the other hand, this low-level to high-level training design explicitly reveals how a high-level concept can be explained by several low-level words.

The Attribute Encoder Module is built on top of Feng's partitioned interpretable embedding [2]. However, for simplicity, we did not apply the mutual independence loss. Moreover, different from their original choices, we use the Unet encoder to encode shape, and a third-party library[3] to extract dominant colors.

## Evaluations

For quantitative result, training metrics and results are listed in Figure 3. Even though MRR $Pnet_{high-level}$ is much lower than $Enet_{pattern}$ and $Enet_{shape}$, high-level concepts have much more labels than pattern and shape labels.

Furthermore, we did a qualitative analysis of 3 high-level concept explanation (Figure 6) by using *AttributeSwap* (Figure 5) operation.

From the tables, we learned the following about high-level concepts: the party scenario favors saturated colors. From the location table, Indian might loves to wear much more colorful clothes than Germany because India appears 7 out 9 rows of colors. Other interesting discoveries can be drawn if we look at the same location across 3 different tables. For example, from (0,0), it tells a story that the British loves wearing a crimson dress for a party during spring-summer.

## Future Work

The proposed model can explain high-level concepts with low-level details. However, we can only narrow down to a few low-level words but the importance of them to the final high-level concept is still agnostic in the proposed solution. As in the quotes of the introduction section, Roland Barthes was wondering about which four features are signifiers of word smart. It actually asked a deeper question about the combination effects of low-level features. We are excited to research more on how to learn the exact weights of low-level words to high-level concepts in our future work.

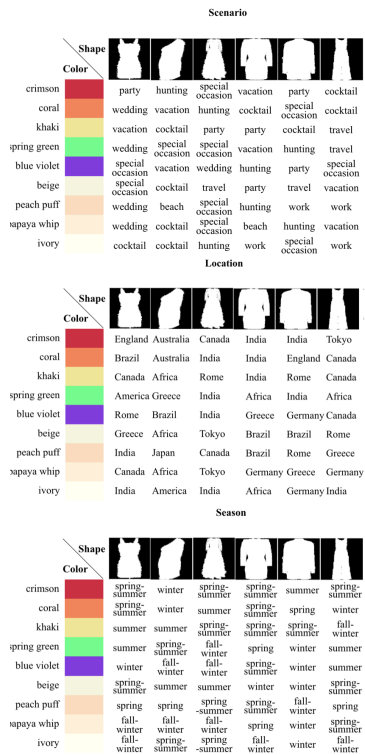

**Figure 6:** Qualitative Result of *AttributeSwap*. The shape in the column represents the target shape that we are going to exchange to, similar for rows. All of the entries share the same pattern embedding. However, labels in the entries is not classification result. It's extracted based on partial order of activation score.